\documentclass[aps,prx,twocolumn,superscriptaddress,longbibliography]{revtex4-2}
\usepackage[colorlinks=true, allcolors=blue]{hyperref}
\usepackage{graphicx}
\usepackage{amsmath}
\usepackage{bbold}
\begin{document}
%\title{Diffuson-mediated crosstalk causes long-range correlated quantum errors}
\title{Long-range correlated quantum errors from diffuson-correlated dephasing}

\author{N.~R.~S. van Venrooij}

\affiliation{Department of Physics and Astronomy, University of Iowa, Iowa City, Iowa 52242, USA}
\affiliation{London Centre for Nanotechnology, University College London, London WC1H 0AH, UK}
\author{M.~E. Flatt\'e}
\affiliation{Department of Physics and Astronomy, University of Iowa, Iowa City, Iowa 52242, USA}\affiliation{Department of Applied Physics and Science Education, Eindhoven University of Technology, Eindhoven 5612 AZ}
\date{\today}
\begin{abstract}
Electrical solid-state qubits are generally initialized, manipulated and read out through diffusive conductors. We show that the disorder-induced low energy modes of such a conductor --- diffusons --- impose spatially correlated dephasing  on any qubits that share it, a mechanism we term \emph{diffuson-correlated dephasing}. Starting from a Keldysh nonlinear sigma model for the disordered conductive region, coupled capacitively to two-level systems, we obtain the  noise kernel in closed form. It decays as the inverse square of the qubit separation and changes sign with the orientation of the pair, so that qubit pairs at equal separation may be correlated or anticorrelated, an unambiguous experimental signature. In two dimensions, the correlated noise a single qubit shares with the rest of the array, in the worst-case scenario, grows logarithmically with array size rather than converging. The sufficient condition for fault tolerance under correlated noise is therefore violated, so standard threshold theorems no longer guarantee protection against this channel for  large qubit arrays. As this dephasing originates from the soft-mode content of the environment rather than on any specific gate detail, we expect correlated dephasing channels of this kind wherever solid-state qubits or sensors share a gapless conductive medium.

\end{abstract}
\maketitle

\section{Introduction}\label{Introduction}

Electrically-addressed solid-state qubit architectures are major candidates for a scalable fault-tolerant quantum computer through the promise of high-fidelity qubits and the industrial manufacturing capability of the semiconductor industry~\cite{Zwanenburg2013, Veldhorst2017, Nickl2026}.
The leading platforms for solid-state qubits fall into several families. Single-atom spin qubits formed by P or As doping in Si or Ge host materials have demonstrated high fidelities and can be fabricated with atomic precision by scanning tunnelling lithography (STL)~\cite{Kane1998, Fuechsle2012, Stock2020} or ion implantation \cite{Morello2020}. Spin qubits can also be made using  physical qubits such as a planar quantum dots (often in composite), leveraging CMOS manufacturing technology ~\cite{LossDiVincenzo1998, Veldhorst2015, Yoneda2018, Nickl2026}. Charge qubits encode information on the charge location directly, such as shared between two localized regions, which can be single-atom or quantum dot-based~\cite{Hayashi2003, Hollenberg2004,
Gorman2005}. 
These architectures share the common feature that the qubits are addressed through extended metallic or degenerately doped regions which  behave as diffusive metals due to inherent disorder~\cite{Shamim2017, Nickl2026}.

Atomistic approaches that aim to understand the microscopic behaviour of these systems, addressing single or few physical qubits embedded in a gate, typically use large-scale numerical simulations and incorporate disorder as a static perturbation to a clean crystalline Hamiltonian~\cite{Klimeck2007, Cifuentes2024}. Systems-level descriptions instead aim to capture large ensembles of interacting qubits and tend to model environment-induced noise phenomenologically, with disorder as one of its sources~\cite{Clemens2004, Cheng2025}. Foundational work\cite{Anderson1958, Abrahams1979, LeeRamakrishnan1985, AltshulerAronov1985} on disordered systems and diffusive metals, however, has shown that even weakly disordered systems behave very differently from their clean counterparts: disorder generates soft collective modes --- diffusons --- which do not have an analogue in the clean limit and dominate the low-frequency response. Treating disorder as a static perturbation to a clean Hamiltonian fails to include these modes. A phenomenological noise model can accommodate the correlations produced by the diffusons but the spatial structure must be postulated rather than derived. 

Here we present a microscopic theory of a long-range correlated dephasing of qubits generated by coupling to the diffuson, a mechanism we term \emph{diffuson-correlated dephasing}. Correlations of this kind are consequential; fault-tolerance thresholds\cite{NgPreskill} are ordinarily quoted under the assumption that errors on distinct qubits are independent, or decay sufficiently fast with separation. Correlated errors have also been resolved recently experimentally in semiconductor arrays~\cite{RojasArias2023, RojasArias2026scaling}. For two qubits that share a two-dimensional disordered conductor separated by $n$ lattice spacings, with in-plane dipoles at angles $\phi_i,\phi_j$ to the axis joining them, the normalised correlation of their dephasing takes the form
\begin{equation}
    \eta_{ij} 
    \approx -\frac{\cos\big(\phi_i+\phi_j\big)}{n^2}.
\end{equation}
Summing its magnitude over a two-dimensional array, the worst-case bound used in threshold theorems~\cite{NgPreskill}, gives a noise strength that grows logarithmically with array size, so that the sufficient condition of these theorems is not met. This has immediate practical scaling implications for qubit architectures that use ground or screening planes \cite{Intel2025}. Such shared layers are likely to become more prevalent as processors scale as delivering control signals to large qubit arrays is a central engineering bottleneck across platforms~\cite{Vandersypen2017, Malinowski2023}.

The result here will be obtained through a field-theoretical derivation employing the Keldysh nonlinear sigma model. The derivation proceeds in four steps. We first formulate the action of the charge qubit on the Keldysh contour, and then, similarly, the action of the disordered environment and of the capacitive coupling between the two. It will then be shown that a spin qubit couples identically to that disordered environment with the same interaction vertex. We work in the Keldysh formalism because it gives direct access to dynamical quantities and permits disorder averaging without recourse to the replica procedure. With these three ingredients in place, we perform a saddle-point analysis of the disordered environment and show that the low-energy effective theory is a non-linear sigma model coupled linearly to the qubit, whose collective diffusive mode --- the \emph{diffuson} --- is the object to which the qubit couples. Finally, having obtained a description of a single qubit coupled to a disordered environment, we introduce a second qubit at separation $R_{ij}$ and determine how the diffuson affects both. The treatment in Sec.~\ref{theory} is, of necessity, highly formal, but readers interested in the main results can focus on Secs.~\ref{theory-intro},\ref{theory-dissipative},\ref{theory-dephasing} for the key results. These will be further discussed and elaborated in Sec.~\ref{discussion} and after.

The final action is similar to a Caldeira-Leggett\cite{CaldeiraLeggett} dissipative action, but derived microscopically instead of postulated phenomenologically. After the main derivation, the discussion  focuses on experimental signatures of the correlated noise channel, the scaling of the noise channel (where we show that the worst-case correlated noise grows without bound with array size), and we finish with a discussion of the extensibility of the method used in the paper. 

\section{Theory}\label{theory}

\subsection{Motivation}\label{theory-intro}
The physical picture preceding the derivation rests on three assumptions:
\begin{enumerate}
    \item A qubit couples capacitively to the electrostatic potential of a nearby disordered conductor.
    \item Only \emph{local} fluctuations of this potential matter: a uniform shift of the gate potential will have an equal effect on all qubits and can therefore be absorbed into calibration. What a qubit responds to are therefore fluctuations about the mean charge density. 
    \item In a disordered conductor a local charge density fluctuation spreads diffusively, through the diffuson. These fluctuations propagate over distances far exceeding the elastic mean free path  $\ell_{\mathrm{el}}$.
\end{enumerate}

These fluctuations reach far because charge is conserved; a local excess of charge cannot vanish, it can only relax by slowly spreading out. A charge imbalance extending over a large region therefore takes a long time to even out, since charge must travel far to do so, and the slowest fluctuations are correspondingly the most spatially extended. Modes of this kind cost vanishingly little energy as their wavelength grows and are referred to throughout this manuscript as soft, massless or low-energy modes. Since dephasing is governed by precisely these slow fluctuations, a single diffuson spreads over a region much larger than $\pi \ell_{\mathrm{el}2}$ and can perturb several qubits sharing the conductor at once. This imposes a \emph{common} stochastic potential rather than a coupling between any particular pair. Such a potential generates spatially correlated errors.

\begin{figure}[thb]
    \centering
    \includegraphics[width=0.8\linewidth, angle=-90, clip]{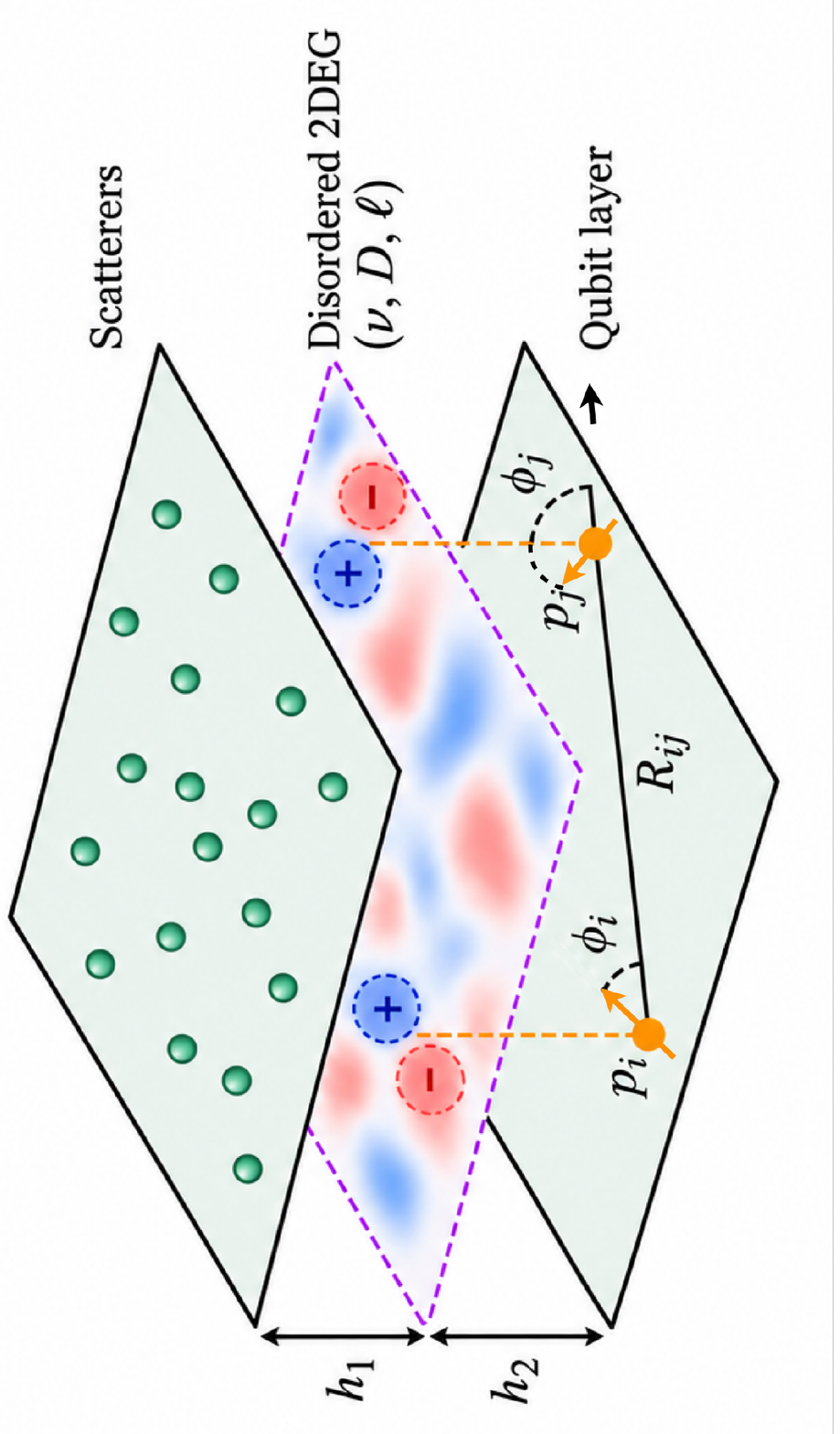}
    \caption{A schematic image of how diffuson-correlated dephasing arises in practice: A random ensemble of scatterers induces a disordered 2DEG that need not be in-plane with the scatterers, hence the distance between the 2DEG and the scatterers is given by $h_1$. The qubits in the qubit layer, separated a distance $h_2$ from the disordered 2DEG, polarize the 2DEG (sketched as the induced dipole above each qubit) and thereby couples to its charge-density fluctuations, which are propagated by the diffuson. Because the diffuson is a soft mode, these correlations decay only as a power law, so every qubit pair in the array is
    correlated.}
    \label{fig:schematic}
\end{figure}

\subsection{Qubits on the Keldysh contour}
We begin by splitting our non-interacting disordered ensemble into two regions: a two-level system (TLS) capacitively coupled to a large, self-averaging, disordered region $B$. Without loss of generality  the TLS that is the focus of this derivation is assumed to be a charge qubit. The TLS Hamiltonian is
\begin{equation}
\begin{aligned}
    H_{\mathrm{TLS}} &= \epsilon_L c^\dagger_L c_L + \epsilon_R c^\dagger_R c_R - \tilde{t}\,(c^\dagger_L c_R + c^\dagger_R c_L)\\
    &= \epsilon_L n_L + \epsilon_R n_R - \tilde{t}\,(c^\dagger_L c_R + c^\dagger_R c_L),
\end{aligned}
\end{equation}
where the operators $c^\dagger_\mu, c_\mu$ ($\mu \in \{L,R\}$) create and annihilate an electron on the left and right orbital of the double-well TLS, $n_\mu = c^\dagger_\mu c_\mu$, $\tilde{t}$ is the tunneling amplitude, and we project onto the one-electron sector by enforcing $n_L + n_R = 1$. Introducing
\begin{equation}
    \tilde\epsilon = \frac{\epsilon_L + \epsilon_R}{2}, \qquad
    \epsilon_z = \frac{\epsilon_L - \epsilon_R}{2},
\end{equation}
we may equivalently write
\begin{equation}
    H_{\mathrm{TLS}} = \tilde\epsilon\,\mathbb{1} + \epsilon_z\,\tau_z - \tilde{t}\,\tau_x,
\end{equation}
where $\mathbb{1}$ is the identity matrix and $\tau_z, \tau_x$ are Pauli matrices in the orbital space $\{L,R\}$.

The corresponding Keldysh action on the closed-time contour is
\begin{widetext}
\begin{equation} 
\begin{aligned}
    S_{\mathrm{TLS}} &= \int_{\mathcal C} dt\, \bar{c}_\mu(t)\,(i\partial_t - H_{\mathrm{TLS}})_{\mu\nu}\,c_\nu(t)\\
    &= \int_{-\infty}^{\infty} dt \, \Big\{ \bar{c}^+_\mu(t)(i\partial_t - H_{\mathrm{TLS}})_{\mu\nu}\,c^+_\nu(t)
    - \bar{c}^-_\mu(t)(i\partial_t - H_{\mathrm{TLS}})_{\mu\nu}\,c^-_\nu(t) \Big\},
\end{aligned}
\end{equation}
\end{widetext}
where $\bar c_\mu, c_\mu$ are the Grassmann coherent-state fields corresponding to $c^\dagger_\mu, c_\mu$, the superscript $\pm$ labels the forward/backward branches of the contour.

Following the Larkin--Ovchinnikov convention, we perform a Keldysh rotation on the fermion fields on the forward and backward branches,
\begin{equation}
\begin{aligned}
    c^1_\mu &= \frac{c^+_\mu + c^-_\mu}{\sqrt{2}}, \qquad
    c^2_\mu = \frac{c^+_\mu - c^-_\mu}{\sqrt{2}},\\
    \bar{c}^1_\mu &= \frac{\bar{c}^+_\mu - \bar{c}^-_\mu}{\sqrt{2}}, \qquad
    \bar{c}^2_\mu = \frac{\bar{c}^+_\mu + \bar{c}^-_\mu}{\sqrt{2}}.
\end{aligned}
\end{equation}
After substitution the action is given by
\begin{equation}
    S_{\mathrm{TLS}} = \int dt\,
    \sum_{ab}\sum_{\mu\nu}\,
    \bar{c}^{a}_\mu(t)\,[G_{\mathrm{TLS}}^{-1}]^{ab}_{\mu\nu}\,c^{b}_\nu(t),
\end{equation}
where $a,b \in \{1,2\}$ are TLS Keldysh indices and $\mu,\nu \in \{L,R\}$ are orbital indices, so that $G_{\mathrm{TLS}}^{-1}$ is a tensor in $\text{Keldysh} \otimes \text{orbital}$. The causality structure lives in the Keldysh indices and is triangular,
\begin{equation}
    [G_{\mathrm{TLS}}^{-1}]^{ab} =
    \begin{pmatrix}
        (G^R)^{-1} & [G^{-1}]^K \\
        0 & (G^A)^{-1}
    \end{pmatrix}^{ab},
\end{equation}
each entry being a $2\times2$ matrix in the orbital indices, with diagonal blocks
\begin{equation}
\begin{aligned}
    (G^R)^{-1} &= i\partial_t - \tilde\epsilon\,\mathbb{1} - \epsilon_z\,\tau_z + \tilde t\,\tau_x + i0^+,\\
    (G^A)^{-1} &= i\partial_t - \tilde\epsilon\,\mathbb{1} - \epsilon_z\,\tau_z + \tilde t\,\tau_x - i0^+,
\end{aligned}
\end{equation}
where the $\pm i0^+$ pole prescription distinguishes the retarded and advanced components. The off-diagonal Keldysh block is not an independent quantity: since $G^K = G^R F - F G^A$, the corresponding entry of the inverse is the pure regulator
\begin{equation}
    [G^{-1}]^K = (G^R)^{-1} F - F (G^A)^{-1} = 2i0^+ F,
\end{equation}
which vanishes as $0^+ \to 0$ and serves only to fix the distribution function $F$, which in equilibrium is $F_\epsilon = 1-2n_F(\epsilon) = \tanh(\epsilon/2T)$.

\subsection{Disordered environment and coupling}

The disordered bath is a free fermionic system coupled to a purely classical, static, spatially disordered potential. Its Keldysh action is
\begin{widetext}
\begin{equation}
    S_{dis} = \int d\mathbf{r}\int dt \sum_{\alpha \beta}\, \Big\{\bar{\psi}^{\alpha}(\mathbf{r},t)\,[G_0^{-1}]^{\alpha \beta}\,\psi^{\beta}(\mathbf{r},t) - V_{dis}(\mathbf{r})\,\bar{\psi}^{\alpha}(\mathbf{r},t)\,\gamma^{cl}_{\alpha \beta}\,\psi^{\beta}(\mathbf{r},t) \Big\},
\end{equation}
\end{widetext}
where $\bar\psi^{\alpha}, \psi^{\alpha}$ ($\alpha \in \{1,2\}$) are the Grassmann fields of the bath electrons in the Keldysh-rotated (LO) basis carrying bath Keldysh indices $\alpha,\beta$, $V_{dis}(\mathbf r)$ is the static disorder potential, and $\gamma^{cl}_{\alpha\beta} = \mathbb{1}_{\alpha\beta}$ is the classical vertex acting in the bath Keldysh space (the disorder is the same on both contour branches, hence purely classical). Throughout, the vertices $\gamma^{cl},\gamma^{q}$ act in whichever Keldysh space their index pair names: $a,b$ for the TLS, $\alpha,\beta$ for the bath. The free inverse Green function $G_0^{-1}$ is triangular in Keldysh space,
\begin{equation}
\begin{aligned}
   \relax [G_0^{-1}]^{\alpha\beta} =&
    \begin{pmatrix}
        (G_0^R)^{-1} & [G_0^{-1}]^K \\
        0 & (G_0^A)^{-1}
    \end{pmatrix}^{\alpha\beta},\\
    \qquad
    (G_0^{R,A})^{-1} =& i\partial_t + \frac{\nabla_\mathbf{r}^2}{2m} + \epsilon_F \pm i0^+,
\end{aligned}
\end{equation}
with the off-diagonal block again the pure regulator fixing the bath distribution.

Charge qubits typically interact with nearby conducting regions through the Coulomb interaction. We approximate the capacitive interaction as an effective short-ranged screened interaction local in space and time; $U(\mathbf{r}-\mathbf{r}',t-t') = U_0\,\delta(\mathbf{r}-\mathbf{r}')\delta(t-t')$. This is appropriate in the regime where the interaction between the qubit and lead can be considered sufficiently screened such that $U(\mathbf q\to 0)\simeq U_0$.

The capacitive coupling between the TLS charge and the bath density, for a charge TLS with charge centers at $\mathbf{R}_L$ and $\mathbf{R}_R$, is described on the Keldysh contour by the source action

\begin{widetext}
\begin{equation}
\begin{aligned}
    S_{c}[\bar{c},c,\bar{\psi},\psi]
    ={}& -U_0 \int_{\mathcal C} dt \int d\mathbf{r}\,
    \Big[
        n_L(t)\,\delta(\mathbf{r}-\mathbf{R}_L)
        +
        n_R(t)\,\delta(\mathbf{r}-\mathbf{R}_R)
    \Big]
    \bar\psi(\mathbf{r},t)\psi(\mathbf{r},t)
    \\
    ={}& -\frac{U_0}{2}
    \int_{\mathcal C} dt \int d\mathbf{r}\,
    \bar c_\mu(t)
    \Big\{
        \mathbb{1}
        \Big[
            \delta(\mathbf{r}-\mathbf{R}_L)
            +
            \delta(\mathbf{r}-\mathbf{R}_R)
        \Big]
        \\
        &\hspace{8.8em}
        +
        \tau_z
        \Big[
            \delta(\mathbf{r}-\mathbf{R}_L)
            -
            \delta(\mathbf{r}-\mathbf{R}_R)
        \Big]
    \Big\}
    c_\nu(t)\,
    \bar\psi(\mathbf{r},t)\psi(\mathbf{r},t).
\end{aligned}
\end{equation}
\end{widetext}

Here we used $n_L+n_R = c^\dagger \mathbb 1\, c$ and $n_L-n_R = c^\dagger\tau_z c$, with $\mathbb 1,\tau_z$ matrices in the orbital space; in the one-electron sector the former is the identity and the latter acts as the Pauli operator $\tau_z$ on $\{|L\rangle,|R\rangle\}$. The part of the action proportional to $\mathbb{1}$ is independent of the TLS state and therefore produces the same shift for both logical states. It may consequently be absorbed into the background potential of the bath. The physically relevant part of the interaction is therefore proportional to $\tau_z$, which describes the exact finite-separation charge-dipole coupling. Applying the Keldysh rotation we obtain the action for the interaction between the qubit and the disordered lead:

\begin{equation}
\begin{aligned}
S_{c}
={}&
-\frac{U_0}{2}
\int dt\,d\mathbf r\;
\Big[\delta(\mathbf r-\mathbf R_L)-\delta(\mathbf r-\mathbf R_R)\Big]
\\
&\times
\Big\{
\rho_z^q(t)\,
\bar\psi^\alpha(\mathbf r,t)\,\gamma^q_{\alpha\beta}\,\psi^\beta(\mathbf r,t)
\\
&\qquad+
\rho_z^{cl}(t)\,
\bar\psi^\alpha(\mathbf r,t)\,\gamma^{cl}_{\alpha\beta}\,\psi^\beta(\mathbf r,t)
\Big\}.
\end{aligned}
\label{S_int}
\end{equation}
with the composite bosonic field terms $\rho_z^{cl}, \rho_z^q$ following the convention

\begin{equation}
\begin{aligned}
2\,\rho_z^q
&= \bar c^+\tau_z c^+ - \bar c^-\tau_z c^-
 = \bar c^{\,a}\tau_z\gamma^{cl}_{ab}c^{\,b},
\\
2\,\rho_z^{cl}
&= \bar c^+\tau_z c^+ + \bar c^-\tau_z c^-
 = \bar c^{\,a}\tau_z\gamma^{q}_{ab}c^{\,b},
\end{aligned}
\end{equation}
where,

\begin{equation}
\begin{aligned}
\gamma^{cl}
&=
\begin{pmatrix}
1&0\\
0&1
\end{pmatrix},
\qquad
\gamma^{q}
=
\begin{pmatrix}
0&1\\
1&0
\end{pmatrix}.
\end{aligned}
\end{equation}
are the classical and quantum vertices in Keldysh space. Note that the $\rho_z$ are TLS scalars (the TLS Keldysh indices $a,b$ are contracted internally), whereas the $\gamma^{cl/q}_{\alpha\beta}$ multiplying the bath bilinears carry the free bath Keldysh indices. 

We take the disorder to be Gaussian and short-range correlated, $\langle V_{dis}(\mathbf{r})V_{dis}(\mathbf{r}')\rangle= (2\pi \nu \tau_{el})^{-1}\delta(\mathbf{r}-\mathbf{r}') $, where $\nu$ is the $d$-dimensional density of states at the Fermi level, considering spinless electrons, and $\tau_{el}$ is the elastic scattering time, which is related to the elastic mean free path through the Fermi velocity $\ell = v_F\tau_{el}$. Throughout the derivation, $\hbar =1$. Disorder averaging in the standard manner generates the usual time-nonlocal four-fermion term, which may be decoupled by a Hubbard--Stratonovich transformation introducing the matrix-valued auxiliary field $Q$ \cite{Kamenev2011}. The qubit interaction acts as an additional bosonic source coupled to the microscopic bath density and therefore remains in the quadratic fermionic kernel. Integrating out the microscopic fields $\bar \psi, \psi$ and re-exponentiating the resulting fluctuation determinant then gives

{\small
\begin{equation}
    \langle Z \rangle_{dis}
    = \int\mathcal{D}[\bar c, c, Q]\,
    \exp\!\Big[\, i S_{TLS} - \frac{\pi\nu}{4\tau_{el}}\operatorname{Tr}Q^2 + \operatorname{Tr}\ln \mathcal{G}^{-1}[\bar c, c, Q] \,\Big],
    \label{Z_dis}
\end{equation}}
where the dressed inverse propagator for the bath fermions is
\begin{equation}
\begin{aligned}
    \big[\mathcal{G}^{-1}[\bar c, c, Q]\big]^{\alpha\beta}
    ={}& [G_0^{-1}]^{\alpha\beta}\,\delta(t-t')\\
    &
    - \frac{U_0}{2} \boldsymbol{\rho}_z\, \Big[\delta(\mathbf r-\mathbf R_L)-\delta(\mathbf r-\mathbf R_R)\Big]\,\delta(t-t')\\
    &
    + \frac{i}{2\tau_{el}}\, Q^{\alpha\beta}_{tt'}(\mathbf{r}),
\end{aligned}
\end{equation}
where
\begin{equation}
    \boldsymbol{\rho}_z \equiv \rho_z^{q} \gamma^q_{\alpha\beta} + \rho_z^{cl}\gamma^{cl}_{\alpha\beta},
\end{equation}
with $\gamma^q_{\alpha\beta}, \gamma^{cl}_{\alpha\beta}$ the quantum and classical vertices in the $\alpha, \beta$ Keldysh subspace.

\subsection{Spin qubits}
\label{sec:spin}

A spin qubit can couple to an electrostatic environment through spin-charge hybridization. Consider a single spin in a harmonic dot subject to a magnetic field gradient, as provided by the micromagnets standard in silicon spin-qubit devices~\cite{Kha2015,Burkard2023}. Its longitudinal component makes the qubit frequency depend on the dot position, and hence renders the qubit charge-sensitive:

\begin{equation}
\begin{aligned}
    H &= -\frac{\hat{\nabla}^2}{2m^*} + \tfrac12 m^*\omega_0^2 \mathbf{r}^2
        + \tfrac12 g\mu_B B_0\,\sigma_z
        + \beta\,(\hat{\mathbf e}\cdot\mathbf r)\,\sigma_z,
    \\
    \beta &= \tfrac12 g\mu_B\,\partial_{\hat{\mathbf e}} B_z .
\end{aligned}
\end{equation}

Since $[H,\sigma_z]=0$, each spin sector is an oscillator whose minimum is
displaced by $\mp\mathbf r_0$, with
\begin{equation}
    \mathbf r_0 = \frac{\beta}{m^*\omega_0^2}\,\hat{\mathbf e},
\end{equation}
so that the logical states are the displaced ground-state orbitals
$\psi_{\uparrow(\downarrow)}(\mathbf r) = \psi_0(\mathbf r \pm \mathbf r_0)$:
the two spin states sit at different positions. The dressed logical subspace is
spanned by the product states
$|\tilde\uparrow\rangle = |\psi_\uparrow\rangle\otimes|\uparrow\rangle$ and
$|\tilde\downarrow\rangle = |\psi_\downarrow\rangle\otimes|\downarrow\rangle$.
In this basis the electronic density operator is diagonal,
\begin{equation}
\begin{aligned}
    \langle \tilde s|\,\hat n(\mathbf r)\otimes\mathbb 1_S\,|\tilde s'\rangle
    &=
    \begin{pmatrix}
        n_\uparrow(\mathbf r) & 0 \\
        0 & n_\downarrow(\mathbf r)
    \end{pmatrix}_{ss'}
    \\
    &= \bar n(\mathbf r)\,\mathbb 1
     + \tfrac12\,\delta n(\mathbf r)\,\sigma_z ,
\end{aligned}
\end{equation}
the off-diagonal elements vanishing identically because $\hat n$ acts only on
the orbital factor while the spin factors are orthogonal. Here
\begin{equation}
\begin{aligned}
    \bar n(\mathbf r) &= \frac{1}{2}\big(|\psi_0(\mathbf r+\mathbf r_0)|^2 + |\psi_0(\mathbf r-\mathbf r_0)|^2\big),\\
    \delta n(\mathbf r) &= |\psi_0(\mathbf r+\mathbf r_0)|^2 - |\psi_0(\mathbf r-\mathbf r_0)|^2 \simeq 2\,\mathbf r_0\cdot\nabla|\psi_0(\mathbf r)|^2 .
\end{aligned}
\end{equation}

In the dressed subspace, the spin-qubit observable  $\sigma_z$ couples to the density difference $\delta n(\mathbf r)$, a dipole density with $\mathbf p=-2\mathbf{r}_0$, which in turn couples capacitively to the disordered conductor exactly as $\tau_z$ does in Eq.~\ref{S_int}. We can thus conclude that in the far field limit, the capacitive coupling of a spin qubit takes the same form as that of a charge qubit, and the derivation below applies to both qubit species. More generally, electrically controlled spin qubits can acquire charge sensitivity through spin--orbital hybridization, including that induced by intrinsic spin--orbit coupling. The resulting qubit vertex is therefore architecture dependent, but the same formalism applies whenever the logical states couple differently to the electrostatic environment.

\subsection{Saddle point and Fluctuations}

Varying the action~\eqref{Z_dis} with respect to the auxiliary field $Q$ yields the saddle-point equation

\begin{equation}
\begin{aligned}
    \frac{\delta S}{\delta Q}
    &=  -\frac{\pi\nu}{2\tau_{el}} Q
    + \frac{i}{2\tau_{el}}\Big( [G^{-1}_0]\,\delta_{tt'}\\
    & \qquad - \frac{U_0}{2} \boldsymbol{\rho}_z\, \Big[\delta(\mathbf r-\mathbf R_L)-\delta(\mathbf r-\mathbf R_R)\Big]\,\delta(t-t')\\
    &\qquad + \frac{i}{2\tau_{el}} Q_{tt'}(\mathbf{r}) \Big)^{-1}\\
    &= 0.
\end{aligned}
\end{equation}
Taking the TLS to shift the saddle only perturbatively, at $O(U_0)$, we drop the composite field and obtain the standard saddle-point equation of the non-interacting disordered electron gas,
\begin{equation}
    \label{Saddle}
    Q = \frac{i}{\pi \nu}\Big( [G^{-1}_0]\,\delta_{tt'} + \frac{i}{2\tau_{el}} Q_{tt'}(\mathbf{r}) \Big)^{-1}.
\end{equation}
This equation is most naturally solved in the energy--momentum basis. For a uniform, stationary solution $Q_{tt'}(\mathbf r) = \Lambda_{t-t'}$, the saddle-point equation is diagonal in energy and momentum. With $G_0^{-1} \to \epsilon - \xi_{\mathbf k}$, where $\xi_{\mathbf k} = k^2/2m - \epsilon_F$, and the inverse evaluated at coinciding points, it reads

\begin{equation}
    \Lambda_\epsilon = \frac{i}{\pi\nu}\sum_{\mathbf k}\Big(\epsilon - \xi_{\mathbf k} + \frac{i}{2\tau_{el}}\Lambda_\epsilon\Big)^{-1}.
\end{equation}

For $\epsilon, \tau_{el}^{-1} \ll \epsilon_F$ the sum may be replaced by $\nu\int d\xi$. The retarded and advanced components are then solved self-consistently by $\Lambda^{R(A)}_\epsilon = \pm1$, since $\sum_{\mathbf k}(\epsilon-\xi_{\mathbf k} \pm i/2\tau_{el})^{-1} = \mp i\pi\nu$. Causality fixes the Keldysh component as $\Lambda^K_\epsilon = \Lambda^R_\epsilon F_\epsilon - F_\epsilon\Lambda^A_\epsilon = 2F_\epsilon$, which in equilibrium is $F_\epsilon = \tanh(\epsilon/2T)$. 

Notice that the stationary configuration obeys a nonlinear constraint $Q^2 = \Lambda^2 = \mathbb{1}$. Any spatially uniform and time-translationally invariant rotation, $Q= \mathcal R^{-1}\Lambda \mathcal R$ where $\mathcal{R} = e^{W/2}$, preserves this constraint and also solves the saddle point equation~ \eqref{Saddle}. Using the cyclic permutation invariance of the trace we may rewrite the trace-log in Eq.~\eqref{Z_dis} and perform a gradient expansion to leading order in slow rotations, valid for $q\ell \ll 1$ and $\omega \tau_{el} \ll 1$, to obtain

\begin{equation}
    iS[\bar c, c, Q] = iS_{TLS} + iS_\sigma[Q] + iS_{\rm c}[\bar c, c, Q],
\end{equation}
where
\begin{equation}
\begin{aligned}
    iS_\sigma ={}& -\frac{\pi\nu}{4}\operatorname{Tr}\big\{D(\nabla_{\mathbf r} Q)^2 - 4\partial_t Q\big\},\\
    iS_{\rm c} ={}& \frac{i\pi\nu U_0}{2}\operatorname{Tr}\big\{\boldsymbol\rho_z\big[\delta(\mathbf r-\mathbf R_L)-\delta(\mathbf r-\mathbf R_R)\big] Q\big\}.
\end{aligned}
\end{equation}
where $D = \frac{v_F^2 \tau_{el}}{d}$, the diffusion constant. This is the familiar Keldysh nonlinear sigma model, but now coupled linearly to the qubit through a composite bosonic field $\boldsymbol{\rho}_z$. Note that we omitted the $O(U_0^2)$-term which in the standard sigma model derivation produces the static compressibility of the electron gas. For the point-like coupling considered here this contribution is a local term and as we are interested in nonlocal effects this term can be safely absorbed in $S_{TLS}$. More detailed derivations of nonlinear sigma models for disordered electronic systems can be found in \cite{Kamenev2011, Altland_Simons_2023, Efetov_1996}. 

It is instructive to now look for the long-range collective excitations of this system. To do so, we allow the Q matrix to fluctuate $Q = \Lambda + \delta Q$. Squaring the fluctuating field gives that $\{ \delta Q, \Lambda \} = 0$, to linear order in $\delta Q$. In order for the fluctuations to obey the nonlinear constraint and hence be an element of the nonlinear sigma model manifold, we can perform a generator expansion of the rotation matrices $\mathcal R$ to fix the functional form of the fluctuations $\delta Q$. Since we are predominantly interested in the effects at thermal equilibrium we may assume the fermionic distribution function $F_\epsilon$ to be spatially uniform and parametrize the saddle as

\begin{equation}
    \Lambda = \mathcal U^{-1} \gamma^{(3)} \mathcal U, \quad     \mathcal U = \mathcal{U}^{-1} =
    \begin{pmatrix}
        1 & F_\epsilon \\
        0 & -1
    \end{pmatrix},
    \quad 
    \gamma^{(3)} = 
    \begin{pmatrix}
        1 & 0 \\
        0 & -1    
    \end{pmatrix}.    
    \label{U}
\end{equation}
Because we are interested in nonuniform rotations of the saddle and the
distribution is uniform, we reparametrize the auxiliary field in the energy basis as
{\small
\begin{equation}
    Q_{\epsilon\epsilon'} \approx \mathcal U^{-1}_\epsilon \big(  \gamma^{(3)} + \tfrac{1}{2}\big[ \gamma^{(3)},W \big] + \tfrac{1}{8}\big[ \big[ \gamma^{(3)},W \big],W \big] + O(W^3) \big) \mathcal U_{\epsilon'}.
    \label{Qexp}
\end{equation}}
$W$ generates nontrivial rotations only if it is Keldysh off-diagonal; it follows that
\begin{equation}
    \delta q_{\epsilon\epsilon'}(\mathbf q) \equiv \frac12\big[\gamma^{(3)},W\big]_{\epsilon\epsilon'}(\mathbf q) =
    \begin{pmatrix}
        0 & d^{cl}_{\epsilon\epsilon'}(\mathbf q) \\
        -d^q_{\epsilon\epsilon'}(\mathbf q) & 0
    \end{pmatrix},
    \label{dq}
\end{equation}
where the bosonic fields $d^{q}, d^{cl}$ are the low-energy modes of the nonlinear sigma model; the diffuson fields, with the energy–momentum representation
$d_{\epsilon\epsilon'}(\mathbf q) = \int d\mathbf r\,dt\,dt'\,d_{tt'}(\mathbf r)\,e^{i\epsilon t-i\epsilon' t'}e^{-i\mathbf q\cdot\mathbf r}$~\cite{Kamenev2011}. Substituting the expansion Eq.~\eqref{Qexp} into $S_\sigma$ gives the free diffuson action

\begin{equation}
    iS^{(2)}_\sigma = \frac{\pi\nu}{2}\int\frac{d\epsilon\,d\omega}{(2\pi)^2}\sum_{\mathbf q}
    d^q_{\epsilon,\epsilon+\omega}(-\mathbf q)\big[Dq^2 - i\omega\big]d^{cl}_{\epsilon+\omega,\epsilon}(\mathbf q),
    \label{Skin}
\end{equation}
with retarded and advanced diffuson propagators
\begin{equation}
\begin{aligned}
    \big\langle d^{cl}_{\epsilon_1\epsilon_2}(\mathbf q)\,d^q_{\epsilon_3\epsilon_4}(-\mathbf q)\big\rangle
    &= -\frac{2}{\pi\nu}\,\delta_{\epsilon_2\epsilon_3}\delta_{\epsilon_1\epsilon_4}\,\mathcal D^R(\mathbf q,\omega),\\
    \big\langle d^{q}_{\epsilon_1\epsilon_2}(\mathbf q)\,d^{cl}_{\epsilon_3\epsilon_4}(-\mathbf q)\big\rangle
    &= -\frac{2}{\pi\nu}\,\delta_{\epsilon_2\epsilon_3}\delta_{\epsilon_1\epsilon_4}\,\mathcal D^A(\mathbf q,\omega),
    \label{propagator}
\end{aligned}
\end{equation}
where
\begin{equation}
    \mathcal D^{R(A)}(\mathbf q,\omega)
    =
    \left[
        Dq^2\mp i\omega
    \right]^{-1}, 
    \qquad 
    \omega = \epsilon_1-\epsilon_2
\end{equation}
In the energy–momentum representation the kernel is the inverse Green function of the diffusion equation $\partial_t n = D\nabla^2 n$. The fluctuating fields $d^{cl,q}$ therefore describe the diffusive spreading of a charge-density fluctuation. In field-theoretical terms, the diffuson is a Goldstone mode of the low energy effective field theory describing the disordered conductor.

The coupling action $S_c$ may be rewritten using cyclicity of the trace, $\operatorname{Tr}\{\boldsymbol{\rho}_z\,\mathcal U^{-1}\delta q\,\mathcal{U}\} = \operatorname{Tr}\{\mathcal{U}\boldsymbol{\rho}_z\,\mathcal U^{-1}\,\delta q\}$, where 
\begin{equation}
\begin{aligned}
    \mathcal{U}_{\epsilon'} \boldsymbol{\rho}_z\,\mathcal{U}^{-1}_{\epsilon} &= \mathcal{U}_{\epsilon'}\big(\rho_z^q\gamma^q_{\alpha\beta} + \rho_z^{cl} \gamma^{cl}_{\alpha\beta}\big)\mathcal{U}^{-1}_{\epsilon} 
    \equiv \rho_z^q\tilde\gamma^q_{\alpha\beta} + \rho_z^{cl} \tilde\gamma^{cl}_{\alpha\beta},
\end{aligned}
\end{equation}
with the rotated Keldysh vertices
\begin{equation}
\begin{aligned}
    \tilde\gamma^{q} ={}& \mathcal{U}_{\epsilon'}\gamma^{q}\mathcal{U}^{-1}_{\epsilon} =
    \begin{pmatrix}
        F_{\epsilon'} & -1 + F_{\epsilon'} F_{\epsilon} \\
        -1 & -F_{\epsilon}
    \end{pmatrix},\\
    \tilde\gamma^{cl} ={}& \mathcal{U}_{\epsilon'}\gamma^{cl}\mathcal{U}^{-1}_{\epsilon} =\,
    \begin{pmatrix}
        1 & F_{\epsilon}-F_{\epsilon'} \\
        0 & 1
    \end{pmatrix}.
\end{aligned}
\end{equation}
Fourier transforming and carrying out the Keldysh traces, the quadratic action reads

\begin{widetext}
\begin{equation}
\begin{aligned}
    iS^{(2)} ={}& iS_{TLS} + \int\frac{d\epsilon\, d\omega}{(2\pi)^2}\sum_{\mathbf q}\Big\{ i\pi\nu\,\Gamma(\mathbf q)\Big[\rho_z^q(\omega)\big((1 - F_\epsilon F_{\epsilon+\omega})d^q_{\epsilon,\epsilon+\omega}(-\mathbf q) - d^{cl}_{\epsilon,\epsilon+\omega}(-\mathbf q)\big)\\
    &\hspace{7em}+ \rho_z^{cl}(\omega)\,(F_{\epsilon+\omega}-F_\epsilon)\,d^q_{\epsilon,\epsilon+\omega}(\mathbf -q)\Big]
    + \frac{\pi\nu}{2}\,d^q_{\epsilon,\epsilon+\omega}(-\mathbf q)\big[Dq^2-i\omega\big]d^{cl}_{\epsilon+\omega,\epsilon}(\mathbf q)\Big\}.
\end{aligned}
\end{equation}
\end{widetext}
Adopting a central-coordinate representation where $\mathbf{R}_i = \tfrac12(\mathbf{R}_L+\mathbf{R}_R)$ and $\mathbf{p}_i = (\mathbf{R_L}-\mathbf{R}_R)$, we may write the vertex 

\begin{equation}
    \Gamma(\mathbf{q}) = \frac{U_0}{2} \big[e^{-i\mathbf{q}\cdot\mathbf{R}_L} - e^{-i\mathbf{q}\cdot\mathbf{R}_R}\big]     =
    -i U_0 e^{-i\mathbf q\cdot\mathbf R_i}\,
    \sin\big({\frac{\mathbf q\cdot\mathbf p_i}{2}}\big).
    \label{vertex}
\end{equation}
Here $\Gamma(\mathbf q)$ denotes the generic qubit-density vertex and is valid for both charge and spin qubits discussed above.

\subsection{Dissipative action}\label{theory-dissipative}

\begin{figure}
    \centering
    \includegraphics[width=0.45\linewidth, angle=-90, trim=4cm 0 4cm 0, clip]{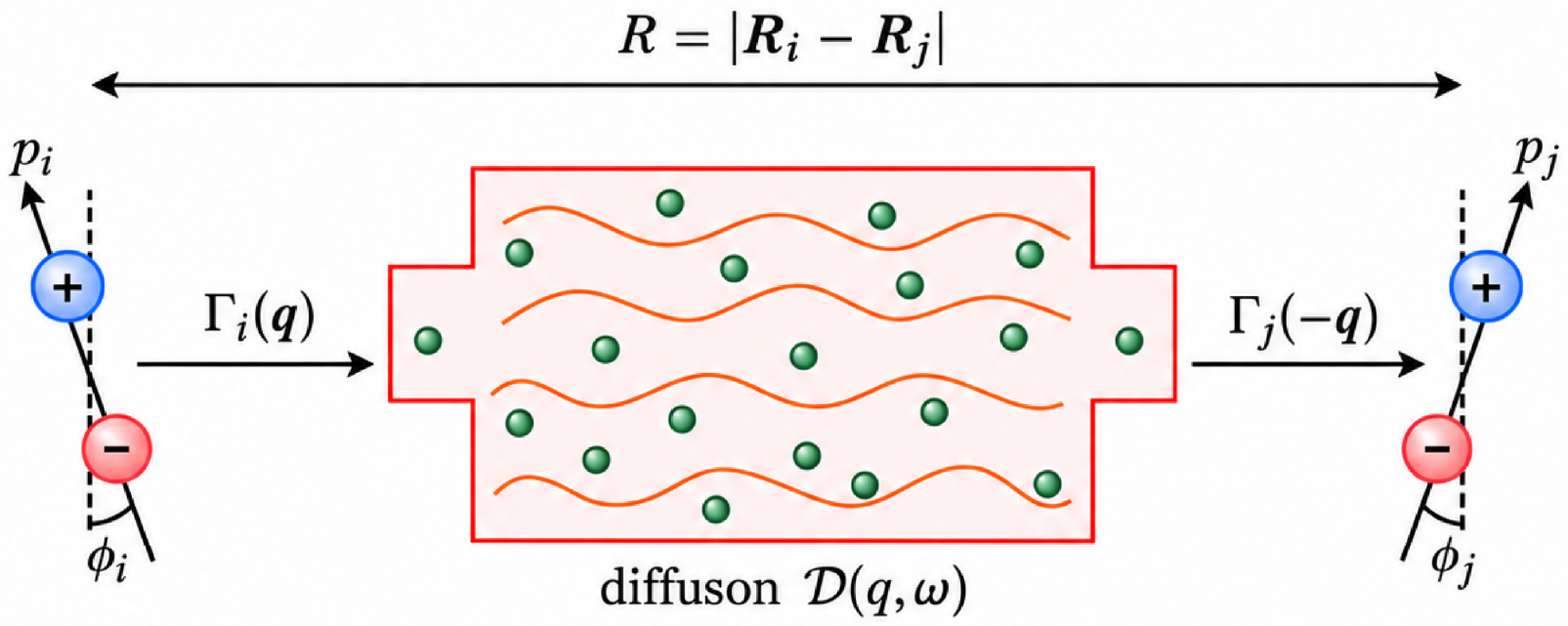}
    \caption{Qubits at sites $\mathbf{R}_i,\mathbf{R}_j$ each couple to a shared disordered conductor through momentum-dependent vertices $\Gamma_i(q),\Gamma_j(-q)$, whose charge-density fluctuations propagate with kernel $\mathcal{D}(q,\omega)$. Sharing the conductor correlates the dephasing of the two qubits, as both are subject to the same fluctuating density field; $\phi_i,\phi_j$ denote the dipole orientations relative to $\mathbf R_{ij}$. In the static limit this is a noise correlator, not a coherent qubit--qubit coupling.}
    \label{fig:crossterm}
\end{figure}

Let us now consider a toy model consisting of two spatially separated TLSs, with central coordinates $\mathbf R_i$ and $\mathbf R_j$, and separation $\mathbf R_{ij} \equiv \mathbf R_i-\mathbf R_j$, with $R_{ij} = |\mathbf R_{ij}|$. Integrating out the diffusive fields $d^q$ and $d^{cl}$  generates three distinct effective interactions: two on-site terms ($i$–$i$ and $j$–$j$) and one cross term. The former cause a renormalization of the TLS on each site through the diffuson vertex and can in principle again be absorbed into the TLS Hamiltonian; the latter is of interest to this work, an illustration of the cross-term is shown in Fig.~\ref{fig:crossterm}. Integrating out the diffusive fields gives the following action for the cross term:
\begin{widetext}
\begin{equation}
\begin{aligned}
    iS_{eff}^{i \leftrightarrow j} ={}&
    -2\pi \nu\int\frac{d\epsilon\, d\omega}{(2\pi)^2}
    \sum_{\mathbf q} \Gamma_i(\mathbf{q})\Gamma_j(\mathbf{-q})
    \Big\{ \big[\rho^q_{z,i}(\omega)\big(1 - F_\epsilon F_{\epsilon+\omega}\big)
    + \rho^{cl}_{z,i}(\omega)\big( F_{\epsilon+\omega} - F_\epsilon\big)\big]\mathcal{D}^{R}(\mathbf{q},\omega)\rho^q_{z,j}(-\omega)\\
    & + \rho^q_{z,i}(\omega)\mathcal{D}^{A}(\mathbf{q},\omega)
    \big[\rho^q_{z,j}(-\omega)\big(1 - F_\epsilon F_{\epsilon+\omega}\big)
    + \rho^{cl}_{z,j}(-\omega)\big(F_\epsilon - F_{\epsilon+\omega}\big)\big]\Big\}.
\end{aligned}
\end{equation}
\end{widetext}
After integrating over the internal fermionic energy $\epsilon$ using integrals $\int d\epsilon\,(1-F_\epsilon F_{\epsilon+\omega}) = 2\omega\coth\frac{\omega}{2T}$ and $\int d\epsilon\,(F_\epsilon - F_{\epsilon+\omega}) = -2\omega$,
the generic effective interaction between two spatially separated TLSs may be written in Keldysh matrix form as

{\small
\begin{equation}
    iS_{eff}^{i\leftrightarrow j}
    =
    4\pi i
    \int\frac{d\omega}{(2\pi)^2}
    \sum_{\mathbf q}
    \Gamma_i(\mathbf q)\Gamma_j(-\mathbf q)\,
    \boldsymbol\varrho_i^{T}(-\omega)\,
    \hat{\Pi}_{nl}(\mathbf q,\omega)\,
    \boldsymbol{\varrho}_j(\omega),
\end{equation}}
where

\begin{equation}
\begin{aligned}
    \boldsymbol{\varrho}_i(\omega)
    &=
    \begin{pmatrix}
        \rho_i^{cl}(\omega)\\
        \rho_i^{q}(\omega)
    \end{pmatrix},\\
    \hat{\Pi}_{nl}(\mathbf q,\omega)
    &=
    i\nu\omega
    \begin{pmatrix}
        0 & -\mathcal D^A \\[2pt]
        \mathcal D^R & 2\coth\!\big(\tfrac{\omega}{2T}\big)\,\mathrm{Re}\,\mathcal D^R
    \end{pmatrix},
\end{aligned}
\label{eq:effective_Keldysh_kernel}
\end{equation}
Notice that the kernel of the effective action, $\Pi_{nl}$, is equivalent to the nonlocal part of the density response of a disordered electron gas. The nonlocal part of the density response is entirely determined by the soft mode of the disorder averaged electronic system: the diffuson. 
The two orderings of the contraction generate the retarded and advanced diffusons, respectively: the noise ($\rho^q\rho^q$) channel collects both orderings and is therefore proportional to the symmetric combination $\mathcal{D}^R + \mathcal{D}^A = 2\,\mathrm{Re}\,\mathcal{D}^R$, while the response ($\rho^{cl}\rho^q$) channel is genuinely retarded (advanced) and carries $\mathcal{D}^R$($\mathcal{D}^A$) alone. Notice that the ($\rho^q\rho^q$) channel is directly related to the Keldysh component of the density-response kernel, and the two channels are tied together by the fluctuation--dissipation theorem,
\begin{equation}
    \Pi_{nl}^{K} = i\nu \omega\coth\tfrac{\omega}{2T}\,\big(\mathcal{D}^R + \mathcal{D}^A\big)
    = \coth\tfrac{\omega}{2T}\,\big(\Pi_{nl}^R - \Pi_{nl}^A\big),
\end{equation}
which follows from $\Pi^{R(A)} = \nu\big(1 \pm i\omega\,\mathcal{D}^{R(A)}\big)$. This confirms that the noise kernel and the dissipative response derive from the same real diffuson spectral weight $\mathrm{Re}\,\mathcal{D}^R$: FDT is satisfied by construction, as it must be for a bilinear coupling to an equilibrium bath. 

The kernel thus correlates TLSs at spatially separated sites through the diffusons; its spatial structure follows from the momentum sum, which admits a closed-form solution
\begin{equation}
\begin{aligned}
&\sum_{\mathbf q}\Gamma_i(\mathbf q)\,\mathcal D^{R(A)}(\mathbf q,\omega)\,\Gamma_j(-\mathbf q)\\
&\quad= \frac{U_0^2}{4}\sum_{s,s'=\pm} s s'\,
\mathcal C_d\big(\mathbf R_{ij}+\tfrac{s\mathbf p_i - s'\mathbf p_j}{2},\omega\big)\\
&\quad\approx -\frac{U_0^2}{4}\,(\mathbf p_i\!\cdot\!\nabla)(\mathbf p_j\!\cdot\!\nabla)\,\mathcal C_d(\mathbf R_{ij},\omega),
\end{aligned}
\end{equation}
where on the last line we applied the dipole approximation, valid for
$|\mathbf p_{i}|, |\mathbf p_{j}| \ll R_{ij}$, and
\begin{equation}
    \mathcal C_d(\mathbf R,\omega)
    = \sum_{\mathbf q}\frac{e^{i\mathbf q\cdot\mathbf R}}{Dq^2 - i\omega}
    =
    \begin{cases}
        \dfrac{e^{-\kappa|R|}}{2D\kappa}, & d=1,\\[2ex]
        \dfrac{K_0(\kappa R)}{2\pi D}, & d=2,
    \end{cases}
\end{equation}
where,

\begin{equation}
    \kappa = \sqrt{\frac{-i\omega}{D}},\quad \operatorname{Re}\kappa > 0 .
\end{equation}

\subsection{Diffuson-correlated dephasing}\label{theory-dephasing}

The effective action of the previous subsection already contains the full result. Since noise models in the quantum-error-correction literature are customarily formulated as stochastic Hamiltonians, we nevertheless recast the noise channel in that language. Defining the noise kernel
\begin{equation}
    \mathcal{K}_{ij}(\omega) \equiv - \frac{\nu U_0^2}{4}\,\omega\coth\tfrac{\omega}{2T}\;\big( \mathbf{p}_i\cdot \nabla \big)\big( \mathbf{p}_j\cdot \nabla \big)\,\mathrm{Re}\,\mathcal C_d(\mathbf{R}_{ij}, \omega),
    \label{Kdef}
\end{equation}
the $\rho^q\rho^q$ term is decoupled by a second Hubbard--Stratonovich transformation, introducing an auxiliary field $\xi_i(\omega)$ with correlation function $\langle \xi_{i,\omega}\,\xi_{j,\omega'}\rangle = 2\pi\,\delta(\omega+\omega')\,\mathcal{K}_{ij}(\omega)$.

In the static limit ($\omega \to 0$), the response channel vanishes whereas the noise channel becomes frequency-independent. This leads to a temporally white but spatially correlated noise field, to which the qubits are subjected through the stochastic Hamiltonian

\begin{equation}
    H(t) = H_{\rm TLS} + \sum_i \xi_i(t)\,\tau_z^{(i)},
    \label{Hstoch}
\end{equation}
where the noise field $\xi$ has covariance

\begin{equation}
    \langle \xi_{i}(t)\,\xi_{j}(t')\rangle \simeq  \mathcal{K}_{ij}(0)\,\delta(t-t'), 
\end{equation}
and the static kernel is written as
\begin{equation}
    \mathcal{K}_{ij}(0) = - \frac{\nu U_0^2T}{2} \,\lim_{\omega\to 0}\big( \mathbf{p}_i\cdot \nabla \big)\big( \mathbf{p}_j\cdot \nabla \big)\,\mathrm{Re}\,\mathcal C_d(\mathbf{R}_{ij}, \omega).
    \label{K0}
\end{equation}

Equations~\eqref{Hstoch}--\eqref{K0} are the central result of this work. They illustrate how in the static limit, qubits that share a disordered conductor are subject to a common, Gaussian dephasing field $\xi$, whose correlations are set by the real-space diffuson propagator $\mathcal{C}_d$ and the dipolar form of the qubit-conductor coupling. None of this is postulated; the long range of the correlations follows from charge conservation, the angular dependence of the noise kernel from the qubit dipole moments $\mathbf{p}_i, \mathbf{p}_j$, and its magnitude depends on the $d$-dimensional density of states $\nu$, the vertex strength between the qubit and the disordered conductor $U_0$, the temperature $T$ and the diffusion constant $D$ (through the diffuson propagator $\mathcal C_d$). More generally, we expect correlated noise of this kind whenever qubits couple through a finite vertex to a soft mode that they share. The spatial structure of the correlated noise and its consequences for fault tolerance are the subject of the next section.

\section{Experimental Signatures and Implications}\label{discussion}

We now elaborate on additional consequences of this theory and propose an experimental signature of the effect.

\subsection{Spatial geometry of the correlation channel}

The most dramatic effect for diffuson correlated dephasing occurs for shared conductors that are two-dimensional on the scale of the qubit separation. Concrete examples are metallic screening or ground planes \cite{Intel2025}, degenerately doped $\delta$-layers that form the reservoirs and leads of STM-patterned Si:P architectures~\cite{Donnelly2025, Shamim2017}, and the extended two-dimensional carrier gas of accumulation-mode heterostructures in which the dots are defined by local depletion~\cite{Borselli_2015}. Other examples may exist depending on the specific architecture. As mentioned earlier, it is expected that as quantum computers scale the wiring problem would likely push architectures to such shared two dimensional conductors. Defining the frequency-dependent diffusion length $L_\omega = \sqrt{D/|\omega|}$, we may expand Eq.~\ref{Kdef} to leading order in $R$ for $R\ll L_\omega$, giving

\begin{equation}
    \mathcal{K}_{ij}(\omega) \approx \frac{\nu U_0^2}{8\pi D} \omega \coth\big(\tfrac{\omega}{2T}\big)\frac{\mathbf{p}_i\cdot \mathbf{p}_j - 2\big( \mathbf{p}_i\cdot \hat{\mathbf{R}} \big)\big( \mathbf{p}_j\cdot \hat{\mathbf{R}} \big)}{R^2}
    \label{eq:Kstatic}
\end{equation}

It is useful to normalise the correlator by its nearest-neighbour amplitude $\bar{\mathcal K} \equiv \frac{\nu U_0^2 p^2}{8\pi D L_q^2}\,\omega\coth\frac{\omega}{2T}$, which for equal dipole magnitudes $|\mathbf p_i| = p$ is independent of orientation. Letting $\phi_i$ denote the angle between $\mathbf p_i$ and $\hat{\mathbf R}_{ij}$ and writing $n = R_{ij}/L_q$, with $L_q$ the qubit pitch, Eq.~\eqref{eq:Kstatic} becomes
\begin{equation}
    \eta_{ij} \equiv \frac{\mathcal K_{ij}}{\bar{\mathcal K}} = -\frac{\cos(\phi_i+\phi_j)}{n^2}.
    \label{eq:eta}
\end{equation}

Vertical confinement in gate-defined devices is an
order of magnitude stiffer than lateral confinement, so the induced dipole
lies predominantly in the plane of the gate; we take $\mathbf p_i$ to be
in-plane throughout. If the system is tuned such that all qubits share a
common dipole orientation, $\phi_i = \phi_j \equiv \phi_{ij}$ and
Eq.~\eqref{eq:eta} reduces to $-\cos(2\phi_{ij})/n^2$. It is thus evident that
the correlation channel is anisotropic and can switch between correlations and
anti-correlations.

The angular structure of Eq.~\eqref{eq:eta} is a parameter-free prediction.
Its sign is fixed by the dipolar vertex, and is therefore independent of the
coupling strength, the diffusion constant and the temperature. This
distinguishes the present mechanism sharply from fluctuator-ensemble models.
We may leverage this structure to suggest an experimental signature that does
not depend on the size of the system. For an equilateral triangle the three
separations are equal, so the radial factor drops out and only the angular
dependence is tested; the three edge orientations differ by $60^\circ$, so
that the arguments of the cosine in Eq.~\eqref{eq:eta} differ by $120^\circ$
and as a result the sum gives
\begin{equation}
    \mathcal{S} \;\equiv\;
    \frac{\mathcal{K}_{12}+\mathcal{K}_{13}+\mathcal{K}_{23}}
         {|\mathcal{K}_{12}|+|\mathcal{K}_{13}|+|\mathcal{K}_{23}|}
    \;=\; 0
    \label{eq:triangle}
\end{equation}
identically, for any shared dipole orientation and any triangle size. By
contrast, any bath whose correlations depend only on $|\mathbf R_{ij}|$ gives
$\mathcal{S}=\pm1$ on the same triangle, irrespective of its decay law. The
statistic is a ratio and is therefore independent of the microscopic coupling
strength: the test requires no estimate of the magnitude of the correlations,
only sufficient statistics to resolve the individual pair correlations.

\begin{figure}
    \centering
    \includegraphics[width=\linewidth]{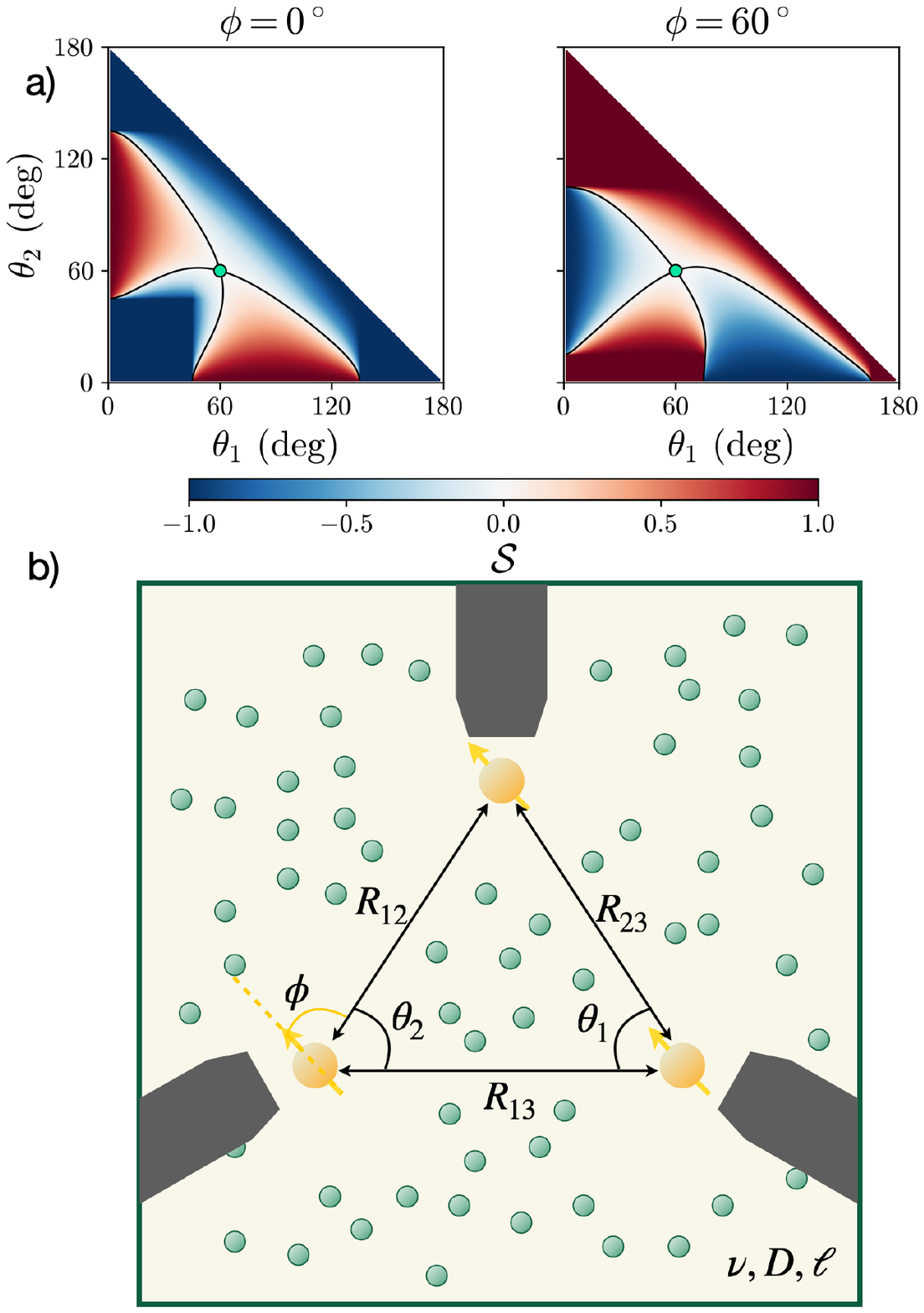}
    \caption{The top two panels (a) show the summed pair correlations $\mathcal{S}$ of three co-aligned qubits with dipole angles $\phi=0^\circ$ and $\phi=60^{\circ}$, in a triangular configuration, ranging from correlated (red) to
    anti-correlated (blue). The bottom panel (b) shows a schematic drawing illustrating what the device should look like: three equidistant qubits (yellow) embedded in a disordered conductor (green) separately addressed by individual qubit gates (grey) for readout. The interior angles $\theta_1,\theta_2$ fix the shape of the triangle up to an overall scale, so the map covers every triangular configuration. For a perfect equilateral triangle the correlations between the qubits cancel exactly. Depending on the dipole    angle there are additional null configurations, indicated by the black zero contours of $\mathcal{S}$. These contours cross at     $(\theta_1,\theta_2) = (60^\circ,60^\circ)$ in both panels, consistent with the equilateral null being independent of $\phi$. }
    
    \label{fig:triangle}
\end{figure}

Here $\mathcal{K}_{ij}$ is proportional to the cross-spectrum of the frequency fluctuations of qubits $i$ and $j$, so that the qubits themselves serve as the probes. The cancellation in Eq.~\eqref{eq:triangle} assumes equal dipole magnitudes with a common orientation and sign, and frequencies low enough that $R \ll L_\omega$. A practical implementation must moreover separate the diffuson contribution from other sources of correlated noise, such as common-mode fluctuations of shared control electronics, which add an isotropic covariance and offset $\mathcal{S}$ from zero. Because the diffuson channel is anisotropic while such backgrounds generically are not, comparing pairs of equal separation but different orientation offers one possible route. We present the triangle geometry as an illustration of how the angular structure of Eq.~\eqref{eq:eta} could serve as a measurement signature, rather than as a complete experimental protocol, and leave its implementation open.

%% =====================================================================

\subsection{Scaling of the noise channel}

The effective theory considered here falls within the same class of Gaussian correlated baths for which fault-tolerance criteria have been formulated in terms of the infrared behaviour of environmental two-point functions~\cite{NgPreskill}. Our method provides the corresponding spatially resolved bath noise kernel microscopically. We may therefore compare its infrared scaling directly with the criteria from these earlier threshold theorems, while leaving a detailed identification of normalisation and operator-ordering conventions outside the scope of the present work. The comparison is particularly relevant because sharper pure-dephasing bounds are often obtained under the assumption that noise on spatially separated qubits is uncorrelated, precisely the assumption violated by the diffuson channel derived here.

The criterion is a statement about the infrared convergence of the noise strength integrated over the array and over the duration of the computation. Under diffusive scaling $R \to \lambda R$, $\omega \to \lambda^{-2}\omega$, the Bessel function $K_0(R\sqrt{-i\omega/D}\,)$ is invariant, while

\begin{equation}
\begin{aligned}
    \big(\mathbf p_i\!\cdot\nabla\big)\big(\mathbf p_j\!\cdot\nabla\big)
    &\rightarrow
    \lambda^{-2}\big(\mathbf p_i\!\cdot\nabla\big)\big(\mathbf p_j\!\cdot\nabla\big),
    \\
    \omega\coth\!\left(\frac{\omega}{2T}\right)
    &\to
    \begin{cases}
        |\omega| \rightarrow \lambda^{-2}|\omega|, & T=0,\\[2ex]
        2T \rightarrow 2T, & |\omega|\ll T,
    \end{cases}
\end{aligned}
\end{equation}
where we have held the dipole moments $\mathbf p_i$ fixed under the rescaling. Consequently, the noise kernel scales as 

\begin{equation}
    \mathcal K(\lambda R,\lambda^{-2}\omega)
    \sim
    \begin{cases}
        \lambda^{-4}\mathcal K(R,\omega), & T=0,\\[3pt]
        \lambda^{-2}\mathcal K(R,\omega), & T\gg |\omega|,
    \end{cases}
\end{equation}
and Fourier transforming gives

\begin{equation}
\begin{aligned}
    \Delta(\lambda R, \lambda^2 t) ={}& \int \frac{d\omega}{2\pi} e^{-i\omega \lambda^2 t}\mathcal{K}(\lambda R, \omega)\\
    ={}& \lambda^{-2}\int \frac{d\omega'}{2\pi} e^{-i\omega' t}\mathcal{K}(\lambda R, \lambda^{-2}\omega')\\
    ={}&
    \begin{cases}
        \lambda^{-6}\Delta(R, t), & T=0,\\[3pt]
        \lambda^{-4}\Delta(R, t), & T\gg |\omega|.
    \end{cases}
\end{aligned}
\end{equation}
Writing the generic scaling relation for a two-point correlation function as $\Delta(\lambda R, \lambda^z t) = \lambda^{-2\delta}\Delta(R,t)$, the real-time noise kernel is infrared convergent when $\tilde D + z < 2\delta$, where $\tilde D$ is the dimensionality of the qubit array. For our $\tilde D = 2$ noise field the condition is met at $T=0$, with $\tilde D + z = 4 < 2\delta = 6$, while it is exactly marginal in the finite temperature static limit $T\gg|\omega|$, where $\tilde D + z = 4 = 2\delta$. 

%The exponent count places the channel at the boundary but cannot say on which side it falls: at marginality the scaling dimension no longer determines whether a quantity saturates or diverges logarithmically, nor with what coefficient. 

In the low-frequency window the noise is white, Eq.~\eqref{Hstoch}, and the noise strength seen by a single qubit reduces to $\sum_j|\mathcal{K}_{ij}(0)|$. Since the normalisation $\bar{\mathcal K}$ is independent of $j$, this factorises as $\sum_j|\mathcal{K}_{ij}(0)| = |\bar{\mathcal K}|\sum_j|\eta_{ij}|$, so that the entire size dependence resides in the parameter-free shape factor and none of it in the microscopic coupling strength.

Evaluating the two sums over a patch of linear size $L$ in the continuum approximation,
{\small
\begin{equation}
\begin{aligned}
    \sum_{j}\eta_{ij} &= 0,\\
    \sum_{j}|\eta_{ij}| &\to \int_{L_q}^{L}\frac{dR}{R}\int_0^{2\pi}d\theta\,|\cos 2\theta|
    = 4\ln\frac{L}{L_q} + c \simeq 2\ln N + c,
\end{aligned}
\label{eq:latticesums}
\end{equation}}
where $N\sim(L/L_q)^2$ is the number of qubits in the patch and $c$ is a constant of order unity. The two sums carry the same scaling dimension and are therefore indistinguishable to the count above, even though one vanishes identically while the other diverges. The coefficient of the divergence is fixed entirely by the angular integral $\int_0^{2\pi}|\cos 2\theta|\,d\theta = 4$ and is as parameter-free as the exponent itself. The vanishing of the signed sum follows from the oddness of $\eta_{ij}$ under $\hat{\mathbf R}_{ij}\to\hat{\mathbf R}_{ij}^{\perp}$, so that any summation region invariant under that interchange cancels term by term. The modulus sum shows that, at the level of the worst case bound, the correlated noise channel diverges in the IR-limit. The absolute scale of the correlated noise is set by the microscopic qubit--environment vertex and is expected to be small for qubits with little charge admixture. What the IR divergence adds, however, is that no infrared suppression protects that small prefactor from growing with array size. The channel may therefore be negligible for small devices and short computations while becoming progressively more relevant as processors grow.

For typical solid-state qubit operating temperatures $k_BT/\hbar$ lies in the sub-GHz to GHz range, whereas error correction and low-frequency dephasing dynamics occur on much lower frequency scales, so the experimentally relevant sector lies firmly in the marginal regime. At finite frequencies, the kernel is cut off beyond the diffusion length $L_\omega$. In real time this means that qubits a distance $R$ apart become correlated only after the diffusion time $R^2/D$, so over a duration $t$ the summed magnitude grows as $2\ln\min(N,Dt/L_q^2)$ and diverges in the joint limit of large arrays and long computations. For $D\sim10$–$100\,\mathrm{cm^2/s}$, $\sqrt{Dt}$ already exceeds $30\,\mu$m within a single microsecond cycle.

We will note that the modulus is the most pessimistic outlook for the effect of the noise kernel on a qubit. It is taken for the same reasons as reported in \cite{NgPreskill}; It assumes nothing about the phase of the qubit's evolution, and therefore forgoes any cancellation against the phase of the bath correlator. Whether the IR divergence of the modulus sum generically survives depends on the qubit dynamics of the relevant architecture, which is beyond the scope of this work. Finally, the quantity relevant to an encoded observable is neither
of these sums. A logical operator supported on a set $S$ accumulates a phase
whose variance is $\sum_{i,j\in S}s_is_j\mathcal{K}_{ij}$, with $s_i=\pm1$ fixed
by the operator; on the compact support of a stabilizer the correlator spans too
narrow a range of orientations for the angular cancellation above to operate, so
the vanishing of the array-wide signed sum should not be read as protection.

\section{Comments on the Theoretical Approach}
At Gaussian order, the present result can equally be obtained by resumming the impurity ladder in the density response and dressing the polarization kernel with the vertices $\Gamma(\mathbf q)$ \cite{LeeRamakrishnan1985, AltshulerAronov1985}; the nonlinear sigma model is not required to derive it. Aside from a stylistic choice, the advantage of the sigma model is that it makes the soft-mode content of the environment explicit. This is  because the sigma model by construction describes the Goldstone manifold of the underlying system. The approach therefore applies for any environment whose low-energy excitations are of a spontaneously broken continuous symmetry. Such materials include disordered electronic systems where the soft (or Goldstone) modes are described by the diffuson and Cooperon, Magnets where the Goldstone modes are magnons, as well as neutral superfluids. Magnetic environments are already used to mediate interactions between spin qubits \cite{Trifunovic2013, Candido_2021, Fukami2021} and we expect a shared magnon field to impose correlated noise of the same structural type as a shared diffuson field. 

%What the response-function route does not supply is the answer to
%which channels are soft in the first place. In the density channel, the
%existence of the diffuson is guaranteed by particle-number conservation but in many situations there exist degenerate soft modes that follow from symmetries of the underlying system. An example of such a soft mode is the Cooperon, which is not the response function of any conserved density, but one that follows from time-reversal symmetry alone, and a symmetry-breaking field gaps it while leaving the diffuson intact \cite{GorkovLarkinKhmelnitskii1979, HikamiLarkinNagaoka1980}. In the sigma model for the disordered electron gas,the Cooperon emerges naturally from the soft manifold (or Goldstone manifold). 

For disordered systems, including superconductors and topological insulators, the Goldstone manifold is fixed by the symmetry class, and the tenfold way ~\cite{AltlandZirnbauer} provides a systematic route to identifying correlated noise channels through selecting the symmetry class to correspond to the material of interest. After determining whether the qubit couples to the respective soft modes of the material, the correlation kernel will follow while the structure of the derivation is unchanged. Where several soft directions coexist the modes mix and the propagators are matrix-valued, which the sigma model handles by construction. Finally, there are regimes in which the Gaussian theory is insufficient altogether: near localization, in low dimensions, or in other critical regimes, fluctuations renormalize the parameters governing the soft modes, and the sigma model is the natural starting point for the RG treatment that then becomes necessary \cite{Wegner1979, EfetovLarkinKhmelnitskii1980}.

During this work, which spanned 2024--2026, various versions of large language models (Claude, Anthropic; ChatGPT, OpenAI) were used as aids in the theoretical analysis. The models were used for distilling key points in literature, checking intermediate steps of derivations and exploring alternative formulations of the derivations. All derivations relied on in this work were further performed and verified, without AI, by the authors.

\section{Outlook}
With minimal assumptions, we identified an emergent correlated dephasing channel for solid-state qubit architectures, carried by the soft mode of the disordered electronic system, the diffuson. The resulting diffuson-correlated dephasing subjects the qubits to long-range correlated noise that decays as the inverse square of the qubit separation. Furthermore, the correlations can change sign depending on the qubit orientations, this angular structure is a direct experimental signature to measure the correlated noise channel in small qubit systems. If the qubits are coupled to a shared two-dimensional disordered conductor, the total noise that one qubit shares with the system was shown to scale logarithmically with the number of qubits in the system - implying an infrared divergence. This means the infrared convergence criterion underlying sufficient conditions underpinning the viability of scaling quantum error correction in quantum computing architectures, in the worst case is not met. This indicates that the effects of correlated noise introduced through soft modes should be accounted for in the design of future architectures. The method extends to other material platforms, where it would serve to extract the functional form of the noise correlation function for direct use in decoders.

The results were obtained through a field theoretical derivation that treated the effects of disorder on both spin and charge qubits microscopically. The field theory is truncated at Gaussian order in the charge density fluctuations, at the same order in perturbations it is possible to include the Cooperon by including time-reversal symmetry. A study of the Cooperon could be of interest for weak-localisation corrections to the dephasing channel. A renormalisation group treatment would be a natural fit for the nonlinear sigma model formalism. The localisation length that such a treatment would provide would technically lead to IR convergence of the diffuson but we don't expect this to counteract the mechanism described in this paper since in practical circumstances the localisation length in two dimensional conductors is exponentially large. The present treatment is restricted to the disorder-averaged kernel and does not resolve mesoscopic effects, which may produce sample-to-sample fluctuations of the correlated noise on the same scale, consistent with the variability already reported in correlation measurements~\cite{RojasArias2023,RojasArias2026scaling}.

A natural extension is to other symmetry classes. In disordered superconductors, the singlet diffuson is gapped, and the relevant question becomes whether the surviving soft modes of the Bogoliubov--de Gennes classes mediate an analogous correlated channel. The mechanism is potentially also of major interest to the quantum sensing community, where entangled shallow NV centers are used for nanoscale sensing~\cite{Rovny2025}. It is plausible that diamond NV-centers couple to the hole gas that forms at the surface of a diamond, and therefore should provide hole-based diffuson-correlated dephasing of the NV-centers. Finally, because collective modes carry weight at long wavelengths, they are not local perturbations on the scale that topological encodings protect against. This makes the mechanism potentially also interesting for topological qubit communities.

\section{Data availability statement}

The code used to generate Fig.~\ref{fig:triangle}(a) is openly available on Zenodo~\cite{triadcode}. All other results follow from the analytical derivations presented in this article.

\section{Acknowledgments}

We thank Andrew Green for extensive discussions on the theoretical framework and for guidance throughout the development of this work. We thank Andrew Fisher for helpful discussions and for pointing us to relevant literature. We would also like to thank Neil Curson and Taylor Stock for discussions and useful information regarding existing experimental platforms. Large language models (Claude, Anthropic) were also used for editing and refining the manuscript text. All text was reviewed and revised by the authors, who take full responsibility for the content. This material (implications for quantum error correction) is based upon work supported by the U.S. Department of Energy Office of Science National Quantum Information Science Research Centers as part of the Q-NEXT center.  The field theoretical foundation of the analysis was supported by the U.S. Department of Energy Office of Basic Energy Sciences through Award No DE-SC0016379.

\bibliographystyle{apsrev4-2}
\bibliography{bibliography}

\end{document}